# Conceptual and Technical Challenges for High Performance Computing

Claude, Tadonki

Mines ParisTech – PSL Research University

Paris, France

Email: claude.tadonki@mines-paristech.fr

## Abstract

High Performance Computing (HPC) aims at providing reasonably fast computing solutions to scientific and real life problems. Many efforts have been made on the way to powerful supercomputers, including generic and customized configurations. The advent of multicore architectures is noticeable in the HPC history, because it has brought the underlying parallel programming concept into common considerations. At a larger scale, there is a keen interest in building or hosting frontline supercomputers; the Top500 ranking is a nice illustration of this (implicit) racing. Supercomputers, as well as ordinary computers, have fallen in price for years while gaining processing power. We clearly see that, what commonly springs up in mind when it comes to HPC is computer capability. However, when going deeper into the topic, especially on large-scale problems, it appears that the processing speed by itself is no longer sufficient. Indeed, the real concern of HPC users is the *time-to-output*. Thus, we need to study each important aspect in the critical path between inputs and outputs. The first step is clearly the *method*, which is a conjunction of *modelling* with specific considerations (hypothesis, simplifications, constraints, to name a few) and a corresponding *algorithm*, which could be *numerical* and/or *non numerical*. Then comes the topic of *programming*, which should yield a skillful mapping of the algorithm onto HPC machines. Based on multicore processors, probably enhanced with acceleration units, current generation of supercomputers is rated to deliver an increasing peak performance, the *Exascale* era being the current horizon. However, getting a high fraction of the available peak performance is more and more difficult. The Design of an efficient code that scales well on a supercomputer is a non-trivial task. Manycore processors are now common, and the scalability issue in this context is crucial. Code optimization requires advanced programming techniques, taking into account the specificities and constraints of the target architecture. Many challenges are to be considered from the standpoint of *efficiency* and expected *performances*. The present chapter will discuss the aforementioned points, interleaved with commented contributions from the literature and our personal views.

**Keywords:** HPC, scalability, manycore, supercomputers, parallelism, computer architecture.

# 1. Introduction

High Performance Computing has been on the spotlight for about two decades, driven by users clamor for more powerful systems and targeting more exciting applications. Significant technical changes have occurred, and noteworthy improvements have been done at various levels, thus pushing the limits of high performance computers. This phenomenon has even changed the rules of scientific discovery. Indeed, large-scale computation is now commonly considered in order to assess if a theory is consistent with experimental results, to question a large collection of data, or to understand a given mechanism through high precision simulations. HPC is thus going hand by hand with cutting-edge research.

At the processor level, the sequential von Neumann execution model has governed the computing landscape for more than a half-century. Thus, the answer for more efficient processing was either a more powerful single-thread processor or an aggregation of cooperative computer systems. Hardware designers have really strived to increase processor capabilities at different levels including *clock speed* (also referred to as *frequency*), instruction level parallelism (ILP), *vector* processing, *memory* size and global *latency*, mass *storage* capacity, and *power* consumption. Regarding parallel computers, they were mainly built by aggregating many standard processors (or machines) with a specific interconnect, thus expensive and very heavy to maintain. Thereby, and also due to the need of a particular skill, parallel computing, which was so far the unique choice for high performance computing, had a very limited effective consideration, despite intensive efforts at the fundamental standpoint. Back to the processor level, chip designers have always strived to stay ahead of Moore's Law, which prescribes that *processor transistors count doubles every two years (Figure 1)*. This was still possible by adding transistors and logic to the standard CPU and increasing clock frequencies, until it becomes exceedingly impractical because of the power wall associated to the increase of processor frequency. Therefore, the idea of multicore processors came up, thus opening the door to the multicore era.

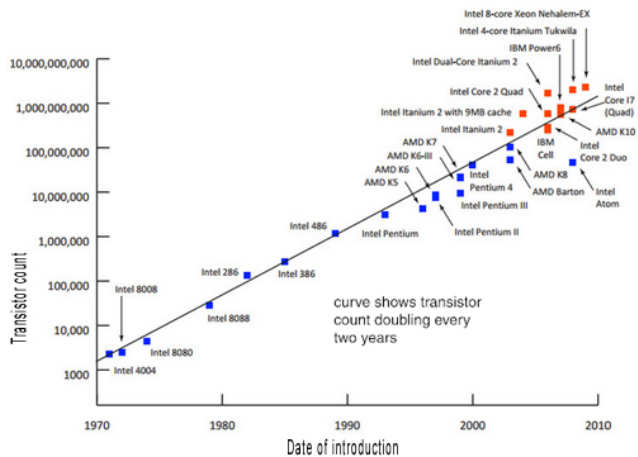

**Figure 1: Microprocessor Transistor Counts 1970-2010 & Moore's Law**

This inflexion point in the evolution of computer systems was the beginning of important technical changes, including the emergence of new hardware devices. With the advent of multicore processors, manufacturers have taken that opportunity to keep providing increasingly powerful processors even to ordinary users, provided that they take the step towards parallel computing. Thereby, the notion of parallelism is extending to a wider audience, and will soon or later become a key item in computer science and engineering curricula. Multicore processors are being actively investigated and manufactured by major computer-processors vendors. At present, most contains between 4 to 16 cores, and a few contain as many as 64 to 80 cores (so-called many-core). In addition, a multi-socket configuration allows getting more cores within the same motherboard. The programmer, in addition to the requirement of an explicit multi-threaded implementation, now has to face more complex memory systems. Indeed, the shared memory available on a multicore processor is typically made of several levels, different packaging and various management policies. Figure 2 displays a basic configuration with the *Nehalem* architecture.

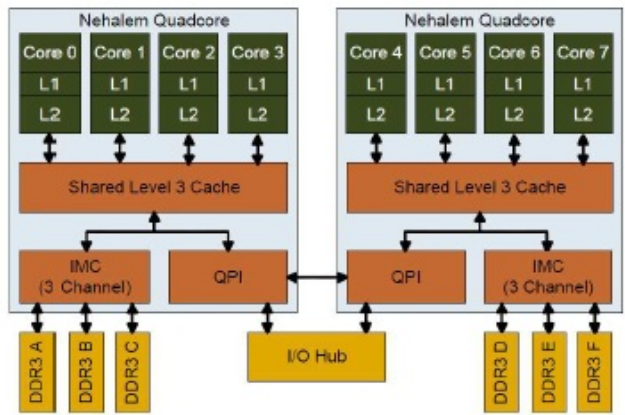

Figure 2: Nehalem Quadcore memory hierarchy

It is important to understand and keep in mind that all *levels of parallelism* need to be skillfully exploited in order to get the highest performance of a given modern (super)computer. The major part of the instruction level parallelism (ILP) is somehow granted by native hardware mechanisms or derived from a suitable instructions scheduling by the compiler. Vector processing, multi-thread computing and multi-process execution need to be managed by the programmer, although some compilers are capable of performing automatic vectorization whenever possible. The necessary skills for achieving an implementation that efficiently combines these specialized programming concepts is likely to stand beyond the reach of ordinary programmers. Valuable tools and libraries exist to assist the programmer in this non-trivial task, but a certain level of expertise is still needed to reach a successful implementation, both from the correctness and the efficiency standpoints. Figure 3 gives an overview of the three aforementioned levels of parallelism.

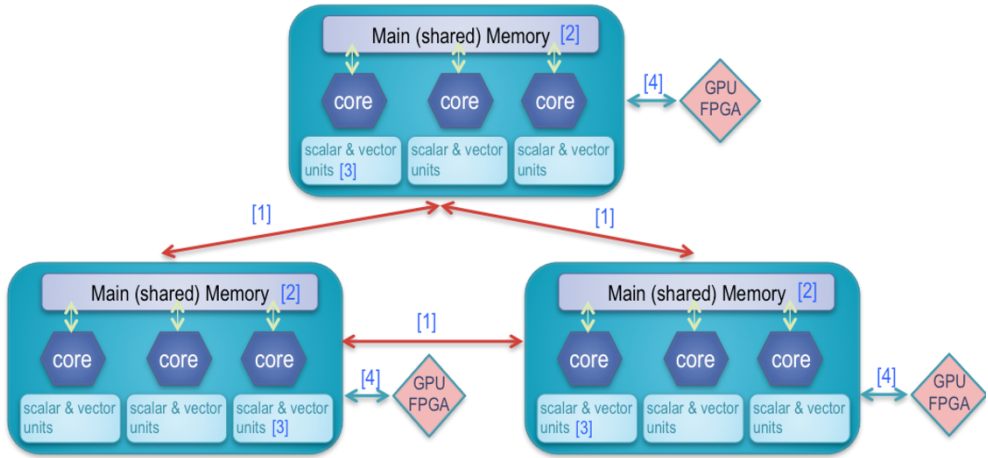

**Figure 3: The three main levels of parallelism**

Figure 3 mentions GPU and FPGA, which are two major accelerators that can be considered for co-processing beside traditional CPUs. Graphic processing unit (usually referred to as GPU) is a specialized microprocessor that offloads and accelerates graphics rendering from the central processor. It was primarily a graphics chip, acting as a fixed-function graphics processor. Gradually, the chip became increasingly programmable and computationally powerful, thereby leading to the GPU. Now, GPU is used jointly with the CPU for general-purpose scientific and engineering applications. The first GPU was designed by NVIDIA, who is still one the leaders of the GPU development, with other companies like Intel and AMD/ATI. The highly parallel structure of modern GPUs makes them very efficient than traditional CPUs for algorithms where processing of large blocks of data can be done in parallel, in addition to classical stream processing applications. This has pushed computer scientist to start thinking about an effective use of GPU to accelerate a wider range of applications, thus leading to the advent of the so-called GPGP (General-Purpose computation on Graphics Processing Units). In GPGPU, a GPU is viewed as a high-performance many-core processor that can be used, under the management of a traditional CPU, to achieve a wide range of computing tasks at a tremendous speed.

Indeed, technical efforts for HPC are noticeable. Advanced techniques are being explored for solving large-scale problems and lot of progresses are made on the programming side. However, number of technical and conceptual challenges remain, some of them being exacerbated by the increasing complexity of current and future HPC systems. Before describing key technical issues, let have a look at the global HPC landscape.

## 2 High-Performance Computing Landscape

Interconnecting a large number of powerful multicore processors (probably accelerated) with a high-speed network is leading to impressive supercomputers. The current horizon is the *Exascale*, which is expected by 2018 (likely 2020 by a linear projection). Supercomputers are doing groundbreaking work that might not be possible without them, and this has changed the rules of science and industry. With computing possibilities running up against the far edge of current technology, researchers are looking for new ways to shrink processors, combine their power, and gather enough energy to make them all work efficiently. Computational capabilities are nowadays an essential part in cutting-edge scientific and engineering experiments. The capability to analyze and predict from huge amount data has incredibly improved with the use of supercomputers. Neuroscientists can evaluate a large number of parameters in parallel to find good models for brain activity; automobile manufacturers can perform more realistic crash simulation to improve safety; astronomers can analyze different regions of the sky in parallel to search for supernovae; nuclear and particle physics are moving beyond common belief with large-scale simulations; search engines can launch parallel search across large-scale clusters of machines and instantly aggregates the results, thus reducing the latency of each request while improving relevance and accuracy; cryptography and computer systems security will benefit from the computation of gigantic prime numbers; researchers in artificial intelligence are trying to use large supercomputers to replicate (or surpass) a high-functioning human's ability to answer questions; social networking services are increasing their pervasiveness through large-scale graph processing, text processing or data mining.

While keep striving to provide faster computers, designers need to contend with power and energy constraints. For decades, computers got faster by increasing their (aggregated) central processor unit. However, high processor frequency means lot of heat. Indeed, The Fujitsu K Computer, for example, has been using US$10 million of electricity per annum to operate. This question of energy is more crucial as computing are being reported to the "Cloud", which is another innovative and affordable way to fulfill the need of high-range computing facilities. Indeed, Cloud computing offers a great alternative on mass storage, software and computing devices. Federating available computing resources, assuming a fast network, is certainly a valuable way to offer a more powerful computing system to the community. Energy, both dissipated and consumed, is also a critical concern, which is subject to active investigation from both the hardware and software standpoints.

From the programming point of view, harvesting hardware advances to rich the level of cutting-edge research expectations is more challenging. Indeed, beside the ambient enthusiasm around the evolution of supercomputers, the way to peak performances is far from straightforward. In addition to algorithmic efforts to express and quantify all levels of parallelism, specific hardware and system considerations have to be taken into account when trying to provide an efficient, robust, and scalable implementation on (heterogeneous) multi-core processors. This has brought an unprecedented level of complexity in program design. Adapting a code for a given architecture or optimize it accordingly requires a complex set of program transformations, each of them addressing one or more aspects (e.g. registers, cache, instruction pipeline, data exchanges) of the target architecture. When the program is complex enough, or when the target architecture is a combination of different processing units (hybrid or accelerated computing), devising highly efficient programs becomes seriously hard. This is the price anyone should be aware of, when it comes to current and future states of high performance computing.

The evolution of supercomputers performance is well depicted in the semi-annual top500 ranking. This has triggered an exciting competition among manufacturers and countries for fastest supercomputers. Being at the frontline in supercomputing infrastructures stands as an evidence of technical and scientific leadership. Alongside the ranking announcements, top500 reports provide a valuable collection of quantitative information for global statistics and trend analysis. Figure 4, for instance, provides a view on the performances evolution (aggregated and extremes) from the beginning of the top500 ranking until November 2017 with a linear extrapolation up to 2020.

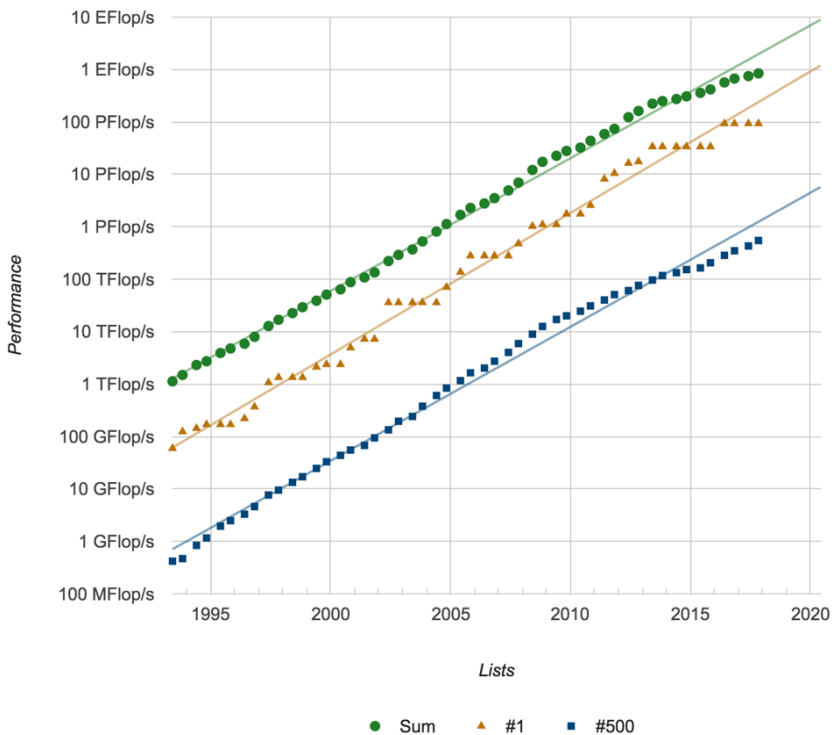

Figure 4: Performance evolution overview from the top500

The petaflops shown up for the first time in June 2008 top500 with the IBM *Roadrunner*, nearly ten years after the reach of the teraflops barrier in June 1997 by Intel *ASCI Red*. The IBM press release (http://www-03.ibm.com/press/us/en/pressrelease/24405.wss) used a few analogies to describe the power of Roadrunner, such as *"The combined computing power of 100,000 of today's fastest laptop computers"*; and, *"It would take the entire population of the earth, - about six billion - each of us working a handheld calculator at the rate of one second per calculation, more than 46 years to do what Roadrunner can do in one day."* From a linear extrapolation, a sustained Exascale performance is expected from 2020. It is amazing to realize that Sunway TaihuLight, the current world fastest supercomputer, is nearly 256 thousand times faster than the top ranked machine of the 1993 top500 edition, the Thinking Machines CM-5/1024. Figure 5 is a snapshot of the top10 machines from the top500 ranking of November 2017.

| # | Site | Manufacturer | Computer | Country | Cores | Rmax [Pflops] | Power [MW] |
|---|---|---|---|---|---|---|---|
| 1 | National Supercomputing Center in Wuxi | NRCPC | Sunway TaihuLight NRCPC Sunway SW26010, 260C 1.45GHz | China | 10,649,600 | 93.0 | 15.4 |
| 2 | National University of Defense Technology | NUDT | Tianhe-2 NUDT TH-IVB-FEP, Xeon 12C 2.2GHz, IntelXeon Phi | China | 3,120,000 | 33.9 | 17.8 |
| 3 | Swiss National Supercomputing Centre (CSCS) | Cray | Piz Daint Cray XC50, Xeon E5 12C 2.6GHz, Aries, NVIDIA Tesla P100 | Switzerland | 361,760 | 19.6 | 2.27 |
| 4 | Japan Agency for Marine-Earth Science and Technology | ExaScaler | Gyoukou ZettaScaler-2.2 HPC System, Xeon 16C 1.3GHz, IB-EDR, PEZY-SC2 700Mhz | Japan | 19,860,000 | 19.1 | 1.35 |
| 5 | Oak Ridge National Laboratory | Cray | Titan Cray XK7, Opteron 16C 2.2GHz, Gemini, NVIDIA K20x | USA | 560,640 | 17.6 | 8.21 |
| 6 | Lawrence Livermore National Laboratory | IBM | Sequoia BlueGene/Q, Power BQC 16C 1.6GHz, Custom | USA | 1,572,864 | 17.2 | 7.89 |
| 7 | Los Alamos NL / Sandia NL | Cray | Trinity Cray XC40, Intel Xeon Phi 7250 68C 1.4GHz, Aries | USA | 979,968 | 14.1 | 3.84 |
| 8 | Lawrence Berkeley National Laboratory | Cray | Cori Cray XC40, Intel Xeons Phi 7250 68C 1.4 GHz, Aries | USA | 622,336 | 14.0 | 3.94 |
| 9 | JCAHPC Joint Center for Advanced HPC | Fujitsu | Oakforest-PACS PRIMERGY CX1640 M1, Intel Xeon Phi 7250 68C 1.4 GHz, OmniPath | Japan | 556,104 | 13.6 | 2.72 |
| 10 | RIKEN Advanced Institute for Computational Science | Fujitsu | K Computer SPARC64 VIIIfx 2.0GHz, Tofu Interconnect | Japan | 795,024 | 10.5 | 12.7 |

**Figure 5: Top ten machines of the November 2012 top500**

Sunway TaihuLight is based on the SW26010 many-core processor (256-core manycore). Each core is clocked at 1.45 GHz and the machine was ranked 16th most energy-efficient supercomputer in the latest Green500, with an efficiency of 6.051 GFlops/watt. This achievement is particularly noticeable. Indeed, in addition of being the first to cross the barrier of 100 PFlops, thus entertaining the hop of hitting the Exascale very soon, it was possible to get a top-ranked performance from both processing and energy standpoints without any accelerator. Nevertheless, the use of accelerators remains a good way to go when it comes to processing-energy efficiency. A nice example was Titan-XK7, the world fastest supercomputer in November 2012 (now 5th). Titan-XK7 is a hybrid supercomputer, means made up by a combination of commodity processors with coprocessors or graphics processing units (GPUs) to form heterogeneous high-performance computing systems. Roadrunner was the world's first hybrid supercomputer, made up with 6,562 dual-core AMD Opteron® chips as well as 12,240 Cell chips (on IBM Model QS22 blade servers). Accelerated computing is prevailing over the use of conventional CPU-based architectures, and is certainly the way to power aware supercomputing. Indeed, as supercomputers are to move beyond the *Petascale* and into the *Exascale*, energy efficiency is becoming a major concern. Note that power consumption, as a metric, was not even mentioned in earlier top500 editions. Now, this aspect has come to the spotlight, and there is a so-called *Green500* project, which aims at providing a ranking of the most energy-efficient supercomputers in the world. We now describe a selection of world-class computing systems.

### A) SUNWAY TAIHULIGHT – SUNWAY MPP

**Sunway TaihuLight**, a system developed by China's National Research Center of Parallel Computer Engineering & Technology (NRCPC), and installed at the National Supercomputing Center in Wuxi, is the current world fastest supercomputer, with a High Performance Linpack (HPL) sustained performance of 93.01 petaflops (over its 125 petaflops peak). Note that this supercomputer is top-ranked for the fourth consecutive time, with more than two times the overall peak of the former number one Tianhe-2. The machine is made up with 40 960 SW26010 many-core processor (256-core manycore), thus a total of 10 649 600 cores. The total amount of available memory is 1280 TB and the (bidirectional) network bandwidth is 16 GB/s. Target applications include *Oil prospecting, life sciences, weather forecast, industrial design, computational cosmology and pharmaceutical research* to name a few. Figure 6 (from http://www.nsccwx.cn/wxcyw/) provides a view of Sunway.

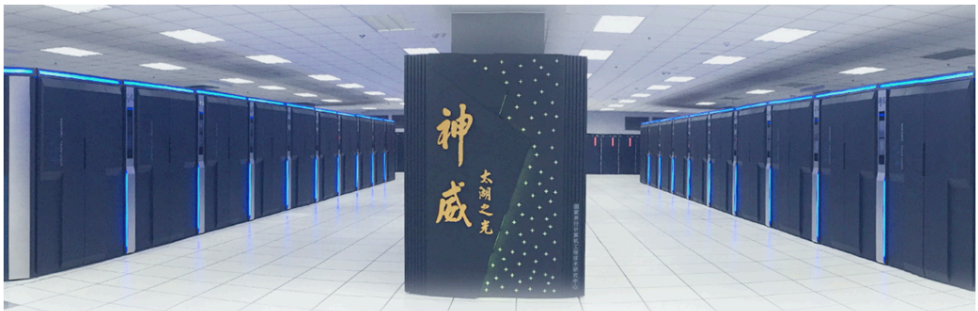

**Figure 6: Sunway TaihuLight Supercomputer**

### B) TIANHE-2 (MILKWAY-2) – TH-IVB-FEP

**Tianhe-2** (**Milky Way-2**), current number two of the latest top500 after being number one just before Sunway, is a system developed by China's National University of Defense Technology (NUDT) and deployed at the National Supercomputer Center in Guangzhou in China. Next to Tianhe-1, Tianhe-2 supercomputer, which showed a High Performance Linpack (HPL) sustained performance of 33.86 petaflops (over its 54.90 petaflops peak), is made up with 16 000 compute nodes, each equipped with two Intel Ivy Bridge Xeon E5-2600 processors and three Xeon Phi coprocessor chips (6x2 + 61x3 = 195 cores), thus making a total of 3 120 000 cores. The total amount of available memory is 1 PB and the (bidirectional) network bandwidth is 10 GB/s. Target applications include *scientific engineering, big data*

*processing, and high throughput computing* to name a few. Figure 7 (from http://en.nscc-gz.cn/) provides a view of Tianhe-2. More technical details about Tianhe-2 can be found in [16].

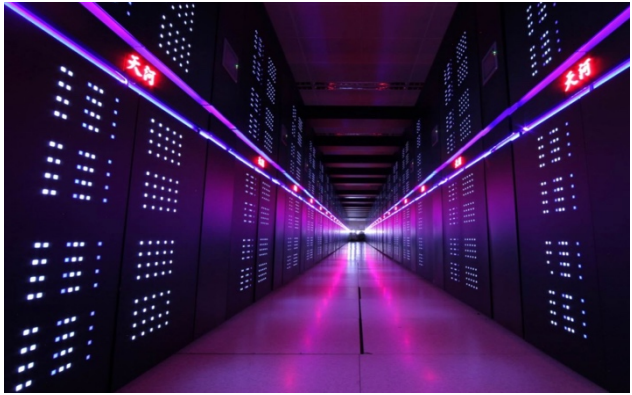

**Figure 7: Tianhe-2 Supercomputer**

### C) PIZ DAINT – CRAY XC50

**Piz Daint**, a Cray XC50 system installed at the Swiss National Supercomputing Centre (CSCS) in Lugano, Switzerland, is the current number three with a High Performance Linpack (HPL) sustained performance of 19.59 petaflops (over its 25.33 petaflops peak), reaffirming its status as the most powerful supercomputer in Europe. Piz Daint a hybrid CPU/GPU supercomputer. Piz Daint supercomputer is made up with hybrid and multicore nodes, Xeon E5-2690v3 12C 2.6GHz and NVIDIA Tesla P100, with a total of 361 760 cores. The total amount of available memory is 437 TB and the average network bandwidth is around 15 GB/s. Target applications include *scientific engineering, big data processing, and high throughput computing* to name a few. Figure 8 (from https://www.cscs.ch/publications/photo-gallery/) provides a view of Piz Daint.

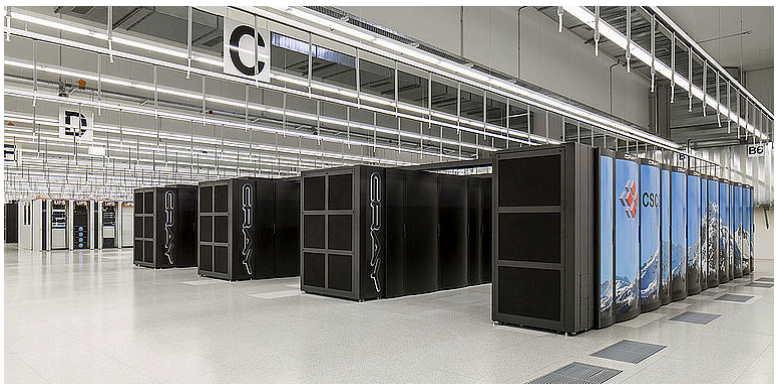

**Figure 8: Piz Daint Supercomputer**

### D) TITAN - CRAY XK7

**Titan – Cray XK7**, also a hybrid CPU/GPU supercomputer manufactured by the Cray Company, was ranked world's fastest supercomputer in the November 2012 top500 ranking. The Cray XK7™, installed at the Department of Energy's Oak Ridge National Laboratory (ORNL / USA), has showed an outstanding 17.59 HLP over a theoretical peak of 27.11 petaflops. The machine is made up with 299,008 16-cores AMD Opteron 6274. This aggregation of CPUs is combined with 261,632 NVIDIA Tesla K20 GPUs. The total memory space available is 710 TB, and the total power consumption is around 8.2 megawatts, which yields a remarkable (rank 2) *performance/power* ratio of 2.14 MFlops/watts. The network is a 3D-torus topology based on the *Gemini interconnect*, which is capable of tens of millions of MPI messages per second with 1.5 microsecond latency and a bandwidth of 20 GB/s for point-to-point transmissions. Figure 9 displays an overview of TITAN.

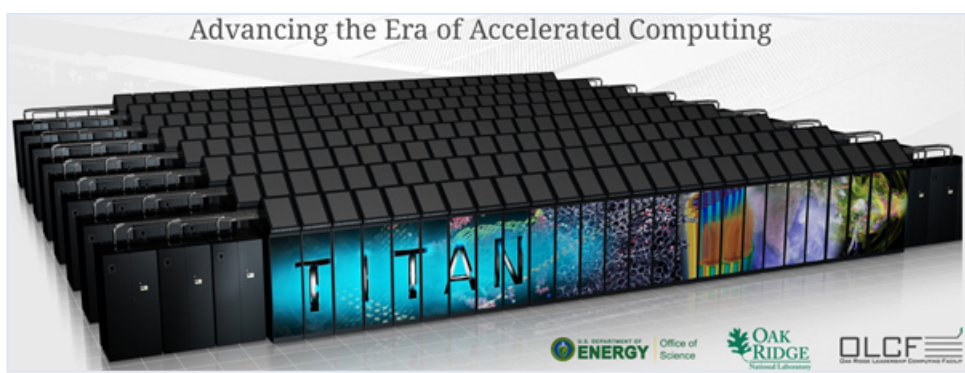

**Figure 9: TITAN Supercomputer**

Among the set of applications that can notably benefit from the tremendous processing speed of Titan, Oak Ridge National Laboratory reported seismological simulations of the entire Earth (suggested by researchers from Princeton University), direct numerical simulation with complex chemistry to understand turbulent combustion, discrete radiation transport calculation, molecular studies, climate change adaptation and mitigation scenario, to name a few. We think that the presence of GPUs should somehow influence the range of potential applications that can be efficiently ported on such machine. A typically suitable application should allow a coarse grain task partitioning with locally interconnected stream processing nodes.

### E) IBM SEQUOIA

**Sequoia** is a world-class IBM BlueGene/Q computer, which was ranked second world's fastest supercomputer in the November 2012 top500 ranking, after being atop in the previous edition. The *Sequoia*, hosted at the Department of Energy's Lawrence Livermore National Laboratory (LLNL / USA), has showed a distinguished 16.32 petaflops HLP over a theoretical peak of 20.14 petaflops. The machine, as mentioned here, is made up with 1,572,864 1.6 GHz cores (16-cores CPUs), with a total memory of 1573 TB. Another attractive strength of Sequoia is its power consumption, which is estimated at 7.9 megawatts, thus making it a good candidate for high-performance computing and high-throughput computing as well. The network is a 5D torus bidirectional optical network with a bandwidth of 5 GB/s and a latency of 2.5 microseconds. Figure 10 illustrates the packaging of Sequoia.

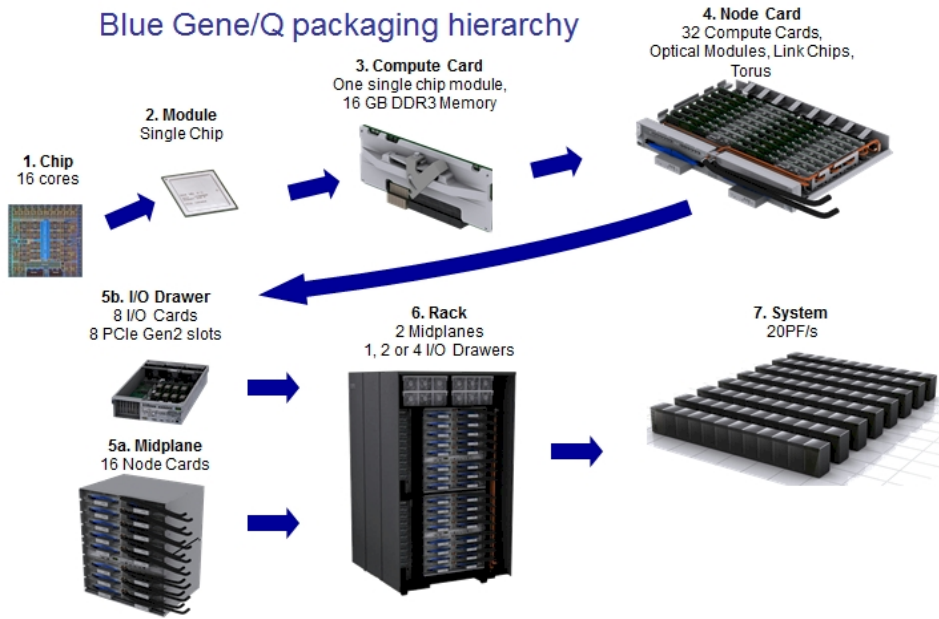

Figure 10: Sequoia packaging

The Sequoia is planned to be eventually devoted almost exclusively to simulations aimed at extending the lifespan of nuclear weapons. However, it flexible interconnect makes it a good choice for (block) stencil computation like the *Lattice Quantum ChromoDynamics* (LQCD) or *Discrete Partial Differential Equation* (DPDE). More classical applications are also considered like *semiconductor and silicone design*, *financial modeling*, *climate and weather* studies. The modest clock rate of each individual core suggests that the machine could be considered for large-scale memory bounded applications. Moreover, the noteworthy low power consumption of the BlueGene/Q makes it clearly adapted for high-throughput computation, with an affordable energy and maintenance cost.

F) Fujitsu K-COMPUTER

K-COMPUTER is a Fujitsu supercomputer, which was ranked third world's fastest supercomputer in the November 2012 top500 ranking, after being atop in the 2011 edition. The K-Computer, hosted at RIKEN Advanced Institute for Computational Science (AICS / Japan), has showed an impressive 10.5 petaflops HLP over a theoretical peak of 11.2 petaflops, thus an excellent processing efficiency. The heart of the K computer consists of

88,128 SPARC64™ VIIIfx 8-cores CPUs, thus a total of 705,024 cores. The overall global memory sums up to 1410 TB. The power consumption is around 12.7 megawatts, which yields a relatively high power per core compared to other machines of the top ten. However, we think that this controversy power consumption is well compensated by the close gap between sustained and peak performances. The K computer's network, called Tofu, uses an innovative structure called "6-dimensional mesh/torus" topology with a total throughput of about 5 GB/s and a microsecond latency for a point-to-point communication between two neighbor nodes. Figure 11 provides a view of K-Computer.

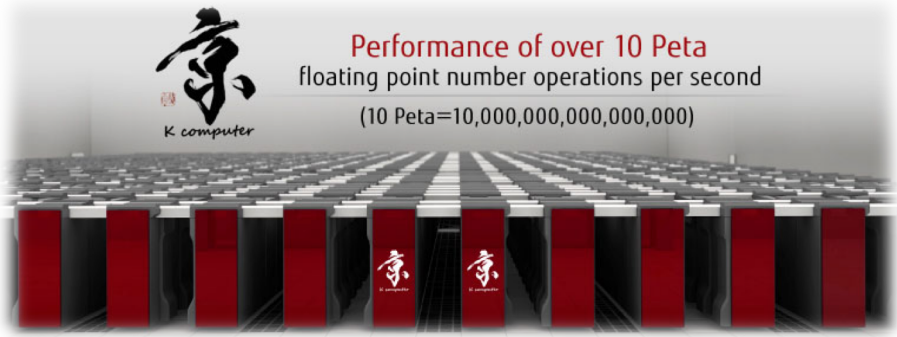

**Figure 11: K-Computer**

The K-Computer has been used on number of successful case studies. First, the machine took the first-place rankings in the 2011 HPC Challenge Awards, which considered various benchmarks aiming at testing different hardware capabilities. In addition, astrophysical N-body simulations of one trillion particles were performed on the full system of the K computer and awarded the 2012 ACM Gordon Bell Prize. The 6-dimensional mesh/torus of the K-computer provides an exceptional communication flexibility, which makes it globally efficient on standard applications. As it uses to be with supercomputers, K-Computer is now open for shared use.

## 3. Major HPC Bottlenecks and Challenges

Let start with some basic quantitative notions related to supercomputers.

a)   Calculating   the overall peak performance

The first thing that comes in mind with a supercomputer is its potential performance, also known (and refers to) as *theoretical peak performance*. This is rough calculation of the overall computing power that the considered computer can offer. The items that are mainly considered are: the total number of cores (regardless of the packaging); the processor clock rate; the length of vector registers (assuming floating point calculations from double precision standpoint); and possibility (or not) of a multiply-add (thus, 2 FP calculations per cycle)

Let consider the case of the **IBM-Sequoia** supercomputer for example. We have

- Total number of cores = 1,572,864
- Processor-core clock rate = 1.6 GHz (i.e. $1,6 \times 10^6$ Hz)
- Each core has Quad FPU (4-wide double precision vector registers)
- One cycle multiply-add feature available

This gives

$1,572,864 \times (1.6 \times 10^6) \times 4 \times 2 = 20.132659 \times 10^{15} \approx 20.14$ **PFlops**

We emphasize on the fact that the time from main memory to the computation units and also the time for interprocessor exchanges are not taken into account. The reader should kept it in mind and be aware that this is where comes the gap between peak and sustained performances. However, intra and extra data routings can be (partially) overlapped with calculations, at the expense of very skillful programming efforts.

**b) Evaluating interprocessor communication**

A supercomputer is composed of a large number or compute nodes that need to exchange data (inputs or intermediate results) in order to achieve the global assigned task. As said above, the time cost for interprocessor communication is roughly seen as an additional time over the pure computing time. For a single data communication, there a setup latency and a transmission time, which gives an estimation of the form $T_c(L) = β + α \times L$. As multiple transfers can occur at the same time, the inverse of the latency (i.e. $1/β$) is sometimes referred in the literature as the number of MPI communication that can be launched within a second. The physical network topology and the current data traffic will determine the effective cost. Indeed, a given point-to-point communication is unlikely to be direct because the communicating nodes might not be directly linked. This is why the physical topology is important, especially in the context of large-scale parallel computers.

### 3.1 About Memory Accesses and Data Transfers

Memory complexity remains a serious challenge both from hardware and software standpoints. Indeed, the part due to memory accesses and data transfers in the sustained performance with common applications is quite significant and even dominant in most cases. At the processor level, this is due to *irregular memory access patterns*, *concurrent accesses*, and *non-uniform memory accesses*. Applications that clearly illustrate this complexity are for example those based on stencil computation (image processing, simulations based on Cartesian space modeling, discrete iterations, computational fluid dynamics, to name a few). In addition to optimizing memory traffic, the programmer now needs to care about cache memories sharing, with a direct consequence on the performance scalability. In a distributed memory context like with distributed memory parallel machines or cloud computing systems, the main bottleneck is the cost of moving data. With accelerators, the main issue is on data exchanges between the accelerator device and the host machine. We now comment each of the aforementioned points.

### 3.1.1 Non-Uniform Memory Access Architectures

Modern CPUs are typically made up with an increasing number of cores in order to deliver a higher peak performance. As the increase of the CPU clock frequency has been somehow frozen because of circuit integration limits and the energy concerns [2], the trend is to provide more and more cores within a single CPU, with a fully shared memory. Nowadays and future supercomputers are just an interconnected aggregation of such nodes [3, 4, 5]. The packaging of a high number of cores within a single chip tends to look like a hardware-connected block of conventional multicores, thus providing a global memory space with a non-uniform access. This Non-Uniform Memory Access (NUMA) configuration is seamless to an ordinary programmer, as there is a unique virtual addressing. Within a NUMA node, the memory system is exactly the same as for an ordinary multicore. Between NUMA nodes, specific links, like the Quick Path Interconnect (QPI), connect local memories together following a specific topology. A memory access in a given NUMA node is said to be local (resp. remote) if the request comes from a core within that (resp. another) NUMA node. This looks like an on-chip distributed memory configuration. Figure 12 displays a typical single-socket NUMA configuration with 4 nodes, while figure 13 (from [6]) illustrates multi-socket cases.

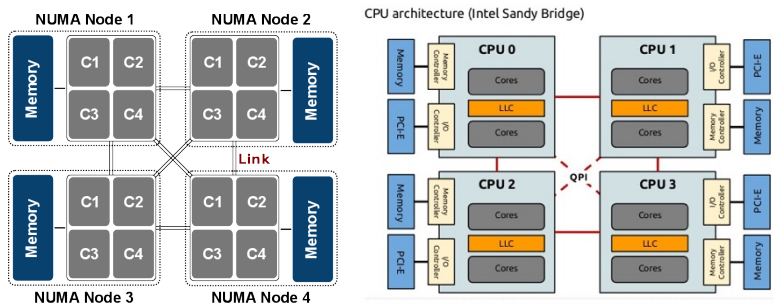

Figure 12: Sample NUMA configurations with 4 nodes

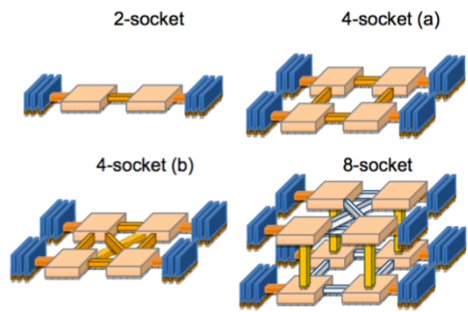

Figure 13: Sample multi-socket NUMA configurations

NUMA configuration was designed to alleviate the bottleneck scenario where all CPU-cores use the same unique bus to access the main shared memory, thereby keeping a high probability of a good scalability over a large number of cores. Unfortunately, good scalability can be obtained only if memory accesses are mostly local. Indeed, remote accesses are more costly by nature and might incur more contention both on the QPI link and within the targeted NUMA node (because local accesses might be carried on at the same time). This potential memory controller saturation or QPI contention is the common culprit of speedup stagnation on NUMA manycores.

Efficient data placement and threads management for better scalability on NUMA systems is a hot topic. Stefan et al. propose a library for parallel programs on NUMA machines, based on array abstraction and memory allocation routines, which allows automatic tuning of data placement and accesses for better scalability [7]. Specific contributions [8, 9, 10, 11, 12] suggest a way to optimize thread and data placement in a NUMA system by combining data locality and thread binding, in order to reduce remote accesses. Lin et al. [13] propose an efficient stencil computation using many-core NUMA architectures, targeting higher performance and portability. Interesting specific cases are studied and reported by C. Tadonki in [14, 15].

### 3.1.2 Data motion in a distributed memory context

#### 3.1.2.1 The Case of Distributed Memory Parallel Machines

We have so far focused on the global processing speed that supercomputers can offer to end-users, with an emphasis on the local efficiency of the computing node and how much is there on the machine. Indeed, a supercomputer is made up with several independent computing nodes, but they need to cooperate and exchange data in order to execute a macroscopic task. What we get from there is the so-called sustained performance, which is most of times far from the theoretical peak. In addition to the gap between sustained and peak performances on a node, there is an additional overhead coming from data exchanges between nodes, which is the main concern of the interconnect efficiency. First note that this aspect is not counted when estimating the peak performance, nor external I/O operations. However, depending on the application, data communication can yield a significant impact on the overall performance, thus breaking the scalability on large-scale supercomputers. The special case of applications involving stencil computation is noteworthy. The Lattice Quantum ChromoDynamics (LQCD), the lattice discretized theory of the strong nuclear force, is a nice example with a gigantic number of sites, each of them having 8 neighbors [16]. The case with a classical graph problem can be found in [26, 27, 28]. When two computing nodes have to exchange data, it is well known that this is better done with a direct communication whenever possible; otherwise a slower multi-hop transfer will take place. The concern here is the mismatch between the virtual topology (from the scheduling) and the physical one (from the target machine). The interconnect of a supercomputer should offer a good flexibility for internode communication. The underlying topology should exhibit either high local degrees or shorter internode distances. Figure 14 outlines a classical interconnect available on supercomputers.

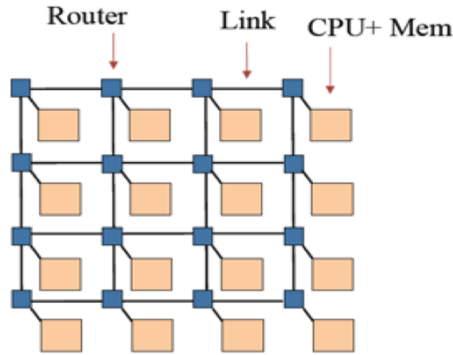

**Figure 14: Typical supercomputer interconnect**

Alongside network topology and bandwidth, communication latency is crucial. The state-of-the-art is around a microsecond, which is acceptable for a point-to-point communication, but less for a multi-hop transfer. Depending on the physical topology and traffic, interprocessor communication might suffer from network congestion, resulting in a significant increase of the sustained latency. Overlapping computation and communication will certainly remain a key ingredient for scalability.

**3.1.2.2 The Case of Cloud Computing Systems with Distant Datacenters**

In the Cloud Computing ecosystem [17, 18], it is common to operate with several distant datacenters, each of them offering different storage capacity and processing speed. This distributed computation, both from task and data standpoints, needs to be skillfully scheduled to achieve an acceptable efficiency, especially in the context where non-locality and heterogeneity apply. The most critical point here is data migration [19, 20]. To achieve good performance and scalability in a Cloud environment with geographically distributed datacenters, data migration should be prevented at the best (through *tasks migration* instead, *processing-migration overlap*, *dataflow optimization*, …) together with an efficient load balancing strategy. Considering a given scheduling methodology, the input workflow might be partitioned into subtasks that are assigned to different datacenters following some of the aforementioned efficiency concerns. Afterwards, processing a subtask might require data migration or replication, which can yield a significant slowdown, especially with a huge amount of data (the case with *big data* applications for instance). An efficient data placement strategy is thus needed in order to yield a more scalable system. This is important for the users as they are under the pay-as-you-go rule, and for the providers too who need to optimize their services and resources pooling.

**3.1.3 Data Exchanges with Accelerators**

In addition to the absolute performance and scalability issues with conventional (multicore) processors, power consumption has quickly become another critical point. The concern is still to compute quite quickly, so as to save energy by reducing the overall running time. The idea that has come in mind to tackle this is the use of accelerators. An accelerator is a specialized unit dedicated to a specific kind of tasks that will be executed with an unbeatable performance. The Graphic Processing Unit (GPU) is one of such devices. At the earlier stage of GPGPU, the main concern was how to efficiently exchange data between the CPU and the GPU. This CPU-to-GPU bottleneck [20], often shirked in some very optimistic reports, has been one of the main hurdles on the GPGPU ascent. Another critical point was the severe slowdown on double precision processing, which is essential in cutting-edge numerical studies. These two issues have been seriously addressed in current generation GPUs, thus making them an effective general purpose computing alternative. In certain applications requiring massive vector operations, this can yield several orders of magnitude higher performance than a conventional CPU. Figure 15 displays an example of processing time improvement of a GPU over a traditional CPU. This example, taken from the NDVIDIA website, reports a benchmark about solving *Navier-Stokes* equations on various grid sizes. Other reported success stories are: a 12x speedup on an orthorectification algorithm and a 41x speedup on the pan sharpening process by Digital Globe; a 3x (resp. 5x) speedup on solving a linear system and a 8x speedup on solving second-order wave equation in MATLAB [21, 22]; a 8x speedup on basic linear algebra subroutines (cuBLAS) [23]; to name a few.

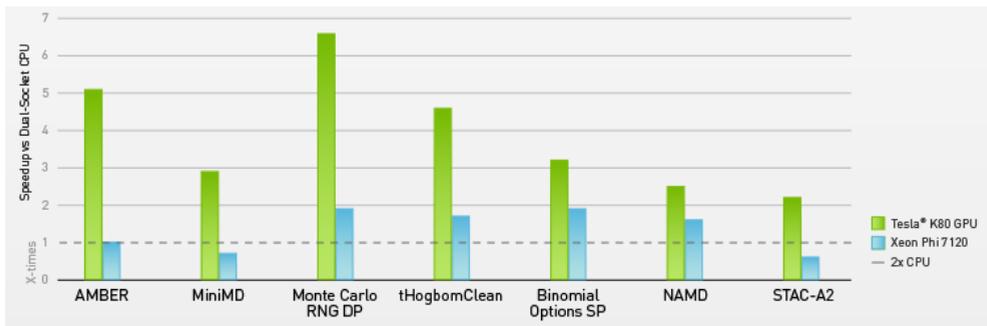

**Figure 15: Illustrative GPU speedups**

The use of GPUs to faster the computation is really coming to the vogue, with the hope of saving energy through shorten execution times. This has motivated the consideration of hybrid CPU/GPU supercomputers and the use of GPU a key device in Cloud computing [24]. Another important point when it comes to parallelism among GPUs is data exchanges, which still need to transit via the referent CPU. This problem is also addressed in current and future generations of GPU, with the aim of having a direct cooperation between the GPUs. Figure 16 illustrates one aspect of the concept via the so-called dynamic parallelism [25].

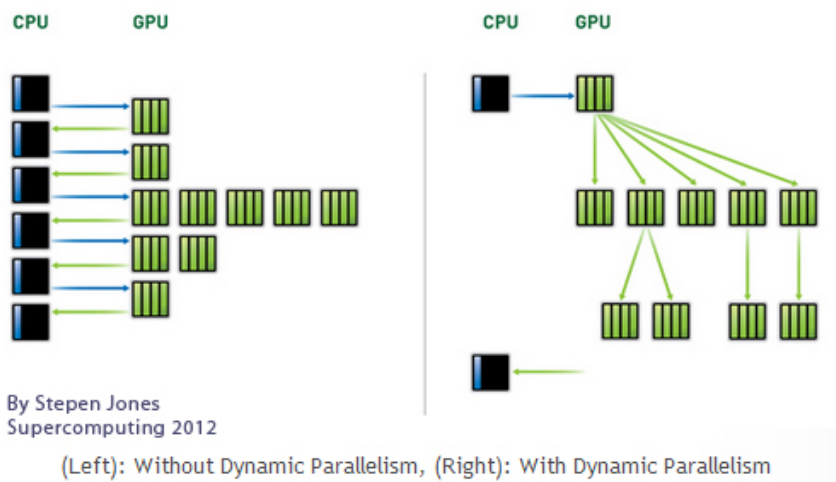

**Figure 16: Dynamic parallelism with GPUs**

## 3.3 Conceptual and Technical Factors Related to Scalability

Processor manufacturers are constantly improving their products by tweaking CPU components and implementing new hardware concepts. The aim is to keep providing increasingly powerful computers for basic issues and large-scale supercomputers for cutting-edge research and engineering. There is a kind of game between progresses and needs, where we iteratively push the limits and try to go beyond. Harvesting computing cycles for science will certainly change the landscape of experimental research and shorten the path to scientific discovery and technical insights.

As we have so far explained, increasing the (aggregated) processor speed raises number of technical challenges that need to be addressed carefully in order to make their benefit clear to the community. Indeed, the gap between the peak performance and the sustained performance is a genuine concern. This is like gross salary and net salary from the employee viewpoint. Users expect supercomputers to be powerful enough for their applications, not in absolute. Thus, getting close to the maximum performance will be a crucial request. From the hardware point of view, this means number of improvement: memory latency at all hierarchy levels should be reduced; opportunity should be given to the programmer to manage memory features as desired; data exchanges between different memory levels should be improved by adding additional buses; the penalty for accessing distant parts of a NUMA memory should be revisited; the set of vector instructions should be soundly extended; network capability should be improved (topology, bandwidth, and latency) in order to lower enough the communication overhead.

The question of heat dissipation and power consumption will sit on top of major concerns. It is possible that, at some points, performance will be sacrificed because of the energy constraint. A typical node of a supercomputer will be made of a traditional multicore processor with several moderate cores, coupled with high-speed accelerator units (mainly GPUs). The idea behind relying on accelerators is that they will be fast enough to significantly reduce the overall execution time, thereby reducing the corresponding heat dissipation. It is important to understand this is a local reasoning, the case of high throughput computation remaining problematic. Indeed, we cannot expect to always compute by spots. Certain kinds of application like simulations, tracking, data assimilation, to name a few, require continuous heavy calculations. The question will be how to keep the benefit of acceleration over a long period of computing time without the punishment of an unacceptable power consumption or hardware failure. Thus, research investigations on the energy efficiency of computing systems will be of a particular interest, both from the hardware side and the programming standpoint. Alongside these efforts, researches on efficient and affordable cooling systems will be also crucial.

Another trend for future innovation, a part from increasing processors horsepower, is the ability to leverage distant power with an increasingly diverse collection of devices. Cloud computing offers a great alternative on mass storage, software and computing devices. Federating available computing resources, assuming sufficiently fast network, is certainly a valuable way to offer a more powerful computing system to the community. The main advantage is that the maintenance cost is mutualized and the users pay only for what they have really consumed. In addition, more related to the *Software as a Service* (SaaS) feature, users instantly benefit from updates, new releases, and new software. There is also an opportunity to share data and key parameters. This approach of federating available resources can be also seen as a way to save power consumption, as it prevents wastage. The topic of *cloud computing* is coming to the vogue and will probably be adopted for major large-scale scientific experiments, assuming non-sensitive data. The challenge for computer scientist is how to efficiently schedule a given set of tasks on the available set of resources in order to serve the request at the user convenience, while taking care of energy.

From the programming point of view, there are number of serious challenges that need to be addressed or remain under deeper investigations. The heterogeneity of current and upcoming supercomputers requires the use of hybrid codes, which is another level of programming complexity. One might think of using (semi-)automatic code generators, thus concentrate on a higher-level abstraction. Programmers will, at certain point, rely on the output of those code generation frameworks, which is not always easy to accept, and otherwise raises a number of practical issues related to *debugging, maintenance, adaptability, tuning, and refactoring*. Figure 17 displays an example of a complex code design framework [3].

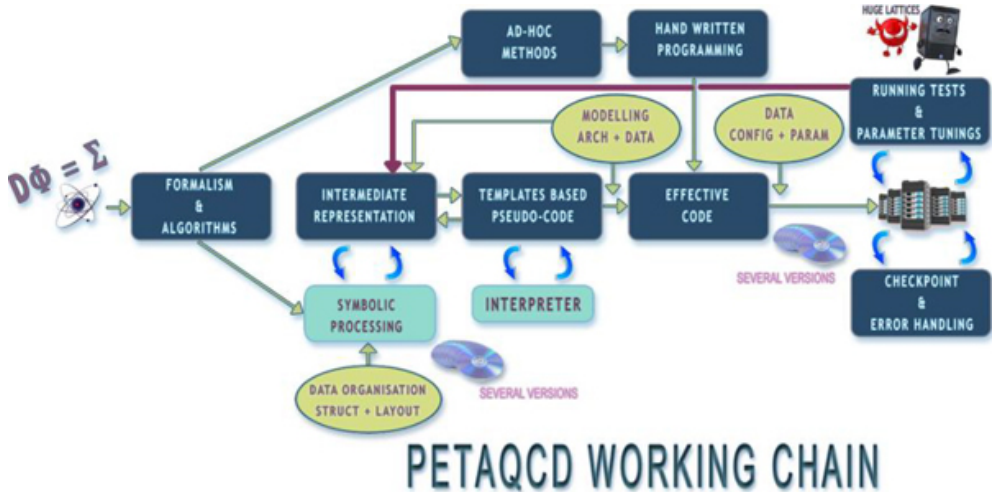
Figure 17: Sample hybrid HPC programming chain

As the number of cores is increasing, with various packaging models, scalability will be an important issue for programmers. Some of the considerations that suited for single-threaded code have to be revised when it comes to multi-threaded version. Data locality is one of them, since the so-called *false sharing* is also caused by an inappropriate locality. Mixing distributed memory model and shared memory model should become a standard.

## 2. Conclusion

High Performance Computing currently stands as a hot topic both for computer scientists and end users. The level of expectations is increasing, motivated by the noticeable technical advances and what is announced at the horizon. Harvesting a high fraction of the available processing power to solve real life problems is a central goal to achieve, as the gap between the theoretical performance and the sustained efficiency is more and more perceptible on modern supercomputers. From the scientific viewpoint, there are number of challenging achievements that are expected in order to come up with efficient and scalable computing solutions. Each involved topic is subject to intensive researches, with significant discoveries that are already effective. However, the connection among these individual advances need to be more investigated. This should be one of the major concerns of future HPC investigations.

Solving large-scale problems in a short period of time using heterogeneous supercomputers is the main concern the high performance computing. We found that combining the advances in *continuous optimization* [24] with suitable mathematical programming formulation of combinatorial problems remains the major approach in operation research. However, there is lack of studies on implementing state-of-the-art optimization methods on modern supercomputers. This is great technical challenge that we should investigate on. The branch-and-bound, for instance, is quite irregular and is likely to exhibit an elusive memory access pattern. Providing a right response to the *load balancing issue* that will certainly show up from a standard scheduling is a challenging task, but very important for *efficiency* and *scalability*. From a fundamental point of view, there is a need to reformulate problems accordingly, with a strong collaboration with people directly involved with real-life applications.

Another interesting topic we which to consider is *automatic code generation* for HPC. Programming current and future supercomputers is becoming more and more difficult, mainly because of their heterogeneity. In addition, obtaining a high fraction of the increasing peak performance is technically hard. One way to obtain an efficient code is to locally optimize each of its critical parts. Another way is to act at the code generation level. Tailoring a code to adapt or achieve the best possible performance on given architecture requires a complex set of program transformations, each designed to satisfy or optimize for one or more aspects (e.g. registers, cache, TLB, and instruction pipeline, data exchanges) of the target system. When the processing code is becoming complex, or when the target architecture is a combination of different processing units (hybrid or accelerated), it becomes very hard to handle the task by hand. Thus, it is highly expected to be able to achieve the necessary code transformations in a systematic way. We should keep investigation this topic, which involves *compilation techniques*, *hardware comprehension*, and *performance prediction*.

# References

[1] Xiang-Ke Liao, Liquan Xiao, Canqun Yang, Yutong Lu, MilkyWay-2 supercomputer: System and application, Frontiers of Computer Science 8(3): 345-356, DOI10.1007/s11704-014-3501-3, 2014.

[2] A. Leite, C. Tadonki, C. Eisenbeis, and A. De Melo, *A fine-grained approach for power consumption analysis and prediction*, Procedia Computer Science (Elsevier), vol. 29, pp. 2260–2271, 2014.

[3] C. Tadonki, *High Performance Computing as a Combination of Machines and Methods and Programming*, Habilitation Thesis, University Paris-Sud, May 2013.

[4] P. Kogge et al., *ExaScale Computing Study: Technology Challenges in Achieving Exascale Systems*, DARPA report, 2008.

[5] E. W. Nagel, D. B. Krner, and M. M. Resch (Eds.), *High Performance Computing in Science and Engineering*, Springer Book Archives, 2013.

[6] Y. Li, I. Pandis, R. Mueller, V. Raman, and G. Lohman, *NUMA-aware algorithms: the case of data shuffling*, http://www.pandis.net/resources/cidr13numashuffling.pdf, 2013.

[7] S. Kaestle, R. Achermann, T. Roscoe, T. Harris, *Shoal: smart allocation and replication of memory for parallel programs*, USENIX Annual Technical Conference, July 8–10, 2015, Santa Clara, CA, USA ISBN 978-1-931971-225,2015.

[8] M. Dashti, A. Fedorova, J. Funston, F. Gaud, R. Lachaize, B. Lepers, V. Quema, M. Roth, *Traffic Management: A Holistic Approach to Memory Placement on NUMA Systems,* ASPLOS'13, March 16–20, 2013, Houston, Texas, USA 2013 ACM 978-1-4503-1870-9/13/03.

[9] B. Lepers, V. Quéma, and A. Fedorova, *Thread and memory placement on NUMA systems: asymmetry matters*, In Proceedings of the 2015 USENIX Conference on Usenix Annual Technical Conference (USENIX ATC '15). USENIX Association, Berkeley, CA, USA, 277-289

[10] R. Lachaize, B. Lepers, and V. Quéma, *MemProf: A memory Profiler for NUMA Multicore Systems*, In USENIX ATC, 2012

[11] A. Collins, T. Harris, M. Cole, C. Fensch, *LIRA: Adaptive Contention-Aware Thread Placement for Parallel Runtime Systems*, In ROSS, 2015

[12] Y. Li, I. Pandis, R. Mueller, V. Raman, and G. Lohman, *NUMA-aware algorithms: the case of data shuffling*, http://www.pandis.net/resources/cidr13numashuffling.pdf, 2013.

[13] P. Lin, Q. Yi, D. Quinlan, C. Liao, Y. Yan, *Automatically Optimizing Stencil Computations on Many-core NUMA Architectures*, International Workshop on Languages and Compilers for Parallel Computing Rochester, NY, United States September 28, 2016 through September 30, 2016.

[14] O. Haggui, C. Tadonki, L. Lacassagne, F. Sayadi, B. Ounid , *Harris Corner Detection on a NUMA Manycore*, Future Generation Computer Systems (DOI: 10.1016/j.future.2018.01.048), 2018.

[15] C. Tadonki, *Scalable NUMA-Aware Wilson-Dirac on Supercomputers*, International Conference on High Performance Computing & Simulation (HPCS 2017), Genoa, Italy, July 17-21, 2017.

[16] Frank Wilczek, *What QCD Tells Us About Nature and Why We Should Listen*, Nuc. Phys. A 663, 320, 2000.

[17] A. Ferreira Leite, A. Boukerche, A. C. Magalhaes Alves de Melo, C. Eisenbeis, C. Tadonki, and C. Ghedini Ralha, *Power-Aware Server Consolidation for Federated Clouds*, Concurrency and Computation: Practice and Experience (CCPE), ISSN: 1532-0626, Wiley Press, New York, USA., 2016

[18] A. Ferreira Leite, V. Alves, G. Nunes Rodrigues, C. Tadonki, C. Eisenbeis, A. C. Magalhaes Alves de Melo, *Dohko: An Autonomic System for Provision, Configuration, and Management of Inter-Cloud Environments based on a Software Product Line Engineering Method*, Cluster Computing Special, 2017.

[19] Y. Samadi, M. Zbakh, and C. Tadonki, *Graph-based Model and Algorithm for Minimizing Big Data Movement in a Cloud Environment*, Int. J. High Performance Computing and Networking, 2018.

[20] Luan Teylo, Ubiratam de Paula, Yuri Frota, Daniel de Oliveira, Lúcia M.A.Drummond, *A hybrid evolutionary algorithm for task scheduling and data assignment of data-intensive scientific workflows on clouds*, Future Generation Computer Systems vol. 17, pp. 1-17, November 2017.

[20] Chris Gregg and Kim Hazelwood, *Where is the Data? Why You Cannot Debate CPU vs. GPU Performance Without the Answer*, International Symposium on Performance Analysis of Systems and Software (ISPASS), Austin, TX. April 2011. http://www.cs.virginia.edu/kim/docs/ispass11.pdf

[21] http://www.mathworks.fr/products/demos/shipping/distcomp/paralleldemo_gpu_backslash.html

[22] http://www.mathworks.fr/company/newsletters/articles/gpu-programming-in-matlab.html

[23] https://developer.nvidia.com/cublas

[24] Frédéric Babonneau, Cesar Beltran, Alain Haurie, Claude Tadonki, Jean-Philippe Vial, *Proximal-ACCPM: a versatile oracle based optimization method,* Optimisation, Econometric and Financial Analysis, vol. 9, pp. 69-92, 2007.

[25] G. Giunta, R. Montella, G. Agrillo, G. Coviello, *A GPGPU Transparent Virtualization Component for High Performance Computing Clouds*, 16th International Euro-Par Conference, Ischia, Italia, August 31 - September 3, 2010.

[26] Stephen Jones, *Inside the Kepler Architecture*, Supercomputing (SC12), Salt Lake City, USA, November 10-16, 2012.

[27] Sanjay Rajopadhye, Tanguy Risset, et Claude Tadonki, *The Algebraic Path Problem Revisited*, European Conference on Parallel Computing, Europar99, Toulouse (France), Lncs Sringer-Verlag, N° 1685, p. 698-707, August 1999.

[28] Giorgos Kollias, Madan Sathe, Olaf Schenk, Ananth Grama, *Fast parallel algorithms for graph similarity and matching*, Journal of Parallel and Distributed Computing, Volume 74, Issue 5, May 2014, Pages 2400-2410.

[29] Laxman Dhulipala, Guy E. Blelloch, Julian Shun, *Theoretically Efficient Parallel Graph Algorithms Can Be Fast and Scalable*, arXiv:1805.05208 , July 2018.